\documentclass[prl,twocolumn]{revtex4}
\usepackage[dvips]{graphicx}

\begin{document}

\bibliographystyle{apsrev}

\title{Coherence Length of Excitons in a Semiconductor Quantum Well}
\author{Hui Zhao}
\author{Sebastian Moehl}
\author{Heinz Kalt}

\affiliation{Institut f\"{u}r Angewandte Physik, Universit\"{a}t
Karlsruhe, D--76128 Karlsruhe, Germany}

\preprint{To be Published in PRL}

\begin{abstract}

We report on the first experimental determination of the coherence
length of excitons in semiconductors using the combination of
spatially resolved photoluminescence with phonon sideband
spectroscopy. The coherence length of excitons in ZnSe quantum
wells is determined to be $300\sim 400$~nm, about $25\sim 30$
times the exciton de Broglie wavelength. With increasing exciton
kinetic energy, the coherence length decreases slowly. The
discrepancy between the coherence lengths measured and calculated
by only considering the acoustic--phonon scattering suggests an
important influence of static disorder.
\end{abstract}

\maketitle

Coherence is an essential intrinsic property of
quantum--mechanical particles. A particle is called coherent if it
propagates like a wave packet with well--defined phases for its
spectral components. Such particles have properties quite
different from classical particles like the ability to show
interference. Constructive interference leads to a macroscopic
coherence of an ensemble of particles. This has profound
consequences as is illustrated by the properties of laser
radiation (coherent ensemble of photons) or by the formation of a
new state of matter in a Bose--Einstein condensate (coherent
ensemble of atoms). The coherence of such an ensemble is destroyed
by phase--relaxing processes of individual ensemble members. The
temporal and spatial scales of phase--destroying processes are
given by the particle coherence time and length, respectively.

The above arguments also hold for excitons, which are the
fundamental quasi--particles of optical excitation in
semiconductors. Recently, coherent control of both ensemble
excitons\cite{l752598} and individual excitons\cite{s2821473} has
been demonstrated, making excitons possible candidates for
quantum--information processing.\cite{s2891906} So it is important
to study the coherence time and length of excitons, which define
an upper temporal and spatial limit for coherent manipulations.
Even for spin manipulations, the temporal or spatial decoherence
can influence spin coherence whenever spin--orbit coupling is
present.\cite{physicstoday5233} In this letter we present the
first experimental determination of the coherence length of
excitons in semiconductors.

Excitonic coherence is temporally and spatially limited due to
interactions within the exciton ensemble and its coupling to its
environment. We can divide these interactions into two classes:
elastic and inelastic scattering. They have essentially different
influences on the excitonic coherence.\cite{bookmitin} For elastic
scattering, the direction of the wave vector changes, but the
energy of the exciton keeps constant. Consequently, the elastic
scattering does not destroy the phase of the exciton. The spatial
distribution of the wavefunction remains independent of time even
after several elastic--scattering events. Thus, the exciton
wavefunction remains coherent.\cite{bookmitin} One can find an
analogue of this coherent propagation in the presence of elastic
scattering in Anderson localization of electrons in
metals.\cite{rmp57287} On the contrary, an inelastic--scattering
event changes the exciton energy and phase and thus destroys the
coherence.

It is nowadays well established to measure the excitonic coherence
time e.g. from the decay of the macroscopic optical polarization
using ultrafast spectroscopy\cite{bookshah} or from the
homogeneous linewidth of exciton luminescence\cite{s27387}.
Methods to investigate spatially coherent phenomena are much less
developed. They require not only a spatial resolution on the order
of the light wavelength\cite{s2932224,s294837} but also a means to
simultaneously test the coherence of the excitons. First attempts
towards this goal were based on time-- and space--resolved
pump--probe experiments.\cite{apl741791,pssb221425} But since the
nonlinear response of the semiconductor was tested, only
high--density regimes beyond the excitonic phase were accessible.
We will show in this letter, that we are able to investigate the
spatial coherence and measure the coherence length of excitons
using spatially resolved phonon sideband spectroscopy in quantum
wells based on the polar semiconductor ZnSe.

The investigations require both specific material properties and a
novel experimental design. The material of choice is a ZnSe/ZnSSe
multiple quantum well for two reasons. Firstly, due to the strong
Fr\"ohlich coupling in polar II--VI quantum structures, one can
optically generate well--defined hot excitons assisted by the
emission of longitudinal optical (LO)--phonons within some
100~fs,\cite{b571390} as shown in Fig.~1. The initial kinetic
\begin{figure}
 \includegraphics[width=5.5cm]{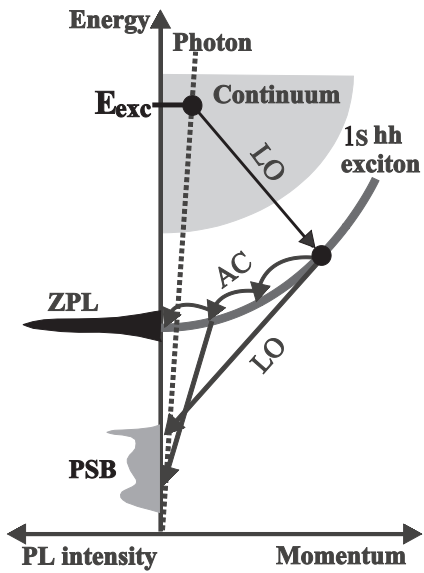}
 \caption{Schematic drawing of the exciton formation assisted by
LO--phonon emission, the subsequent relaxation by acoustic phonon
(AC) emission and the resulting zero--phonon--line (ZPL) and
phonon sideband (PSB).
 }
\end{figure}
energy of the exciton can be well controlled over a range of
30~meV simply by choosing the laser excitation energy,
$E_{\mathrm{exc}}$. This exciton formation mechanism is
drastically different from that of III--V
structures\cite{b65035303}, in which most of the optically
generated electrons and holes relax individually toward their
respective band extrema, and bind to excitons on a time scale of
10~ps.\cite{l713878} Secondly, the strong Fr\"ohlich coupling
induces a pronounced LO--phonon sideband (PSB) beside the
zero--phonon--line (ZPL) in the photoluminescence (PL) spectrum.
After their formation, the hot excitons relax to the band minimum
by phonon emission. The relaxed excitons with nearly zero kinetic
energy can radiatively recombinate, resulting in the ZPL. Since
the photon momentum is very small, this direct coupling to photons
is forbidden for hot excitons due to momentum conservation. This
is the main obstacle of hot--exciton studies in conventional PL
spectroscopy, which exploits the ZPL. For the same reason, the hot
excitons are not observable in pump--probe experiments. However,
the LO--phonon can assist the hot exciton in coupling with a
photon by taking away its momentum (see Fig.~1). The simultaneous
well--defined energy loss to the LO--phonon leads to the
appearance of the PSB. This process is an ideal tool for the {\it
direct} investigation of hot excitons.\cite{bookexciton,b571390}

Figure~2 shows three spectra excited by a continuous--wave
\begin{figure}
 \includegraphics[width=7cm]{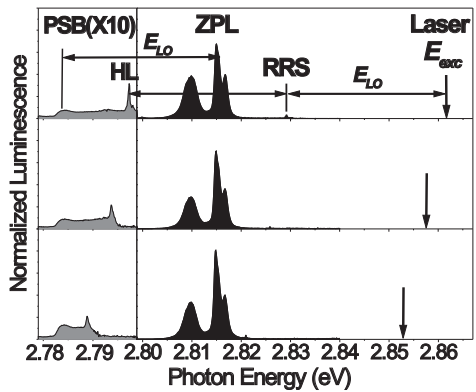}
 \caption{Photoluminescence spectra measured at 7 K. The
 excitation intensity is 1 $\mathrm{kW/cm^{2}}$.
 }
\end{figure}
laser at various photon energies. The PSB is found in the
photon--energy range between one LO--phonon energy
($E_{\mathrm{LO}}$=31.8 meV) below the ZPL and $2E_{\mathrm{LO}}$
below the $E_{\mathrm{exc}}$ (see Fig.~2). The latter energy
defines the upper limit of the PSB, which stays a strict limit
when $E_{\mathrm{exc}}$ is varied. This fact confirms that the PSB
really reflects the hot--exciton distribution and that no
additional luminescence, e.g., related to defects, occurs in this
spectral range. In Fig.~2, a sharp peak (HL) is observed at
$2E_{\mathrm{LO}}$ below $E_{\mathrm{exc}}$. This peak is the main
subject of the present investigation. In the following we will
identify its origin and discuss what we can learn from this peak.

Generally, the peak observed at
$E_{\mathrm{exc}}-nE_{\mathrm{LO}}$ can be induced by hot--exciton
luminescence (HL) and/or resonant Raman scattering
(RRS).\cite{pssb689} There has been a debate in the 1970's on how
to distinguish these two processes.\cite{b9622,b12624} The actual
difference between them is whether a real excitonic population is
involved or not.\cite{b9622} HL is composed of two distinct
one--photon processes, i.e., absorption followed by emission,
while RRS is a single two--photon process.\cite{l67128} In the
case of $n=1$, HL is not possible,\cite{pssb689} thus the peak can
be easily identified as a first order RRS. Indeed, this peak is
observed in Fig.~2 on the high--energy side of the ZPL. In the
cases of $n\geq2$, both HL and RRS are possible. In experiments,
one can distinguish them according to their different
features.\cite{pssb689}. Recently, it has been proven that HL can
be the dominant process at low temperature in several systems,
e.g., II--VI\cite{l67128,b439354}, III--V\cite{apl601615} and
IV--IV\cite{b64085203} semiconductors. In our experiments, we
prove that the sharp peak observed at the upper limit of PSB is
dominated by HL by the following experimental facts. At first,
raising the temperature we observe a pronounced thermal quenching
of this peak (from 800 counts at 10~K to less than 100 counts at
70~K), while the RRS peak keeps unchanged in this temperature
range (about 160 counts). Such different temperature behaviors
clearly distinguish HL and RRS. \cite{pssb689,b64085203} Secondly,
increasing the photon energy of excitation by $E_{\mathrm{LO}}$,
the spectral linewidth of the HL peak becomes 2~$\sim$~3 times
wider. This is also a signature of an HL process.\cite{b439354}
Beside these facts, the different spatial profiles of the HL and
RRS peaks (see below) also provides evidence of their origins.

Having excluded a significant Raman contribution to the HL peak,
we will see that the peak can be exploited to study the individual
coherence of excitons. The peak is spectrally located at
$2E_{\mathrm{LO}}$ below $E_{\mathrm{exc}}$ thus
$1E_{\mathrm{LO}}$ below the initial kinetic energy of the
exciton. So it actually monitors the population of excitons right
after their LO--phonon assisted formation and before their first
inelastic--scattering event. An exciton that has undergone the
first inelastic--scattering event, thus has a different energy,
cannot contribute to the peak. Since an exciton remains coherent
between two inelastic--scattering events, {\it the HL peak at
$2E_{\mathrm {LO}}$ below $E_{\mathrm{exc}}$ monitors the presence
of individual coherent excitons} with a well--defined kinetic
energy $E_{\mathrm{exc}}-E_{\mathrm{LO}}$.

Our experimental setup is a solid immersion lens (SIL)--enhanced
confocal micro--photoluminescence system with a resolution of
450~nm (FWHM). The details of the system have been described
previously.\cite{APL801391} By moving a pinhole in the image plane
of the microscope, we can scan the detection spot with respect to
the excitation spot, thus study the spatial coherence in a rather
direct way. Also, the high collection efficiency of the setup
enables us to access the low--density regime where
exciton--exciton interactions are negligible.

By scanning the pinhole, we measure the spectra of the PSB at
different detection positions with respect to the excitation spot,
as shown in Fig.~3(a). We find the HL peak drops with increasing
\begin{figure}
 \includegraphics[width=7cm]{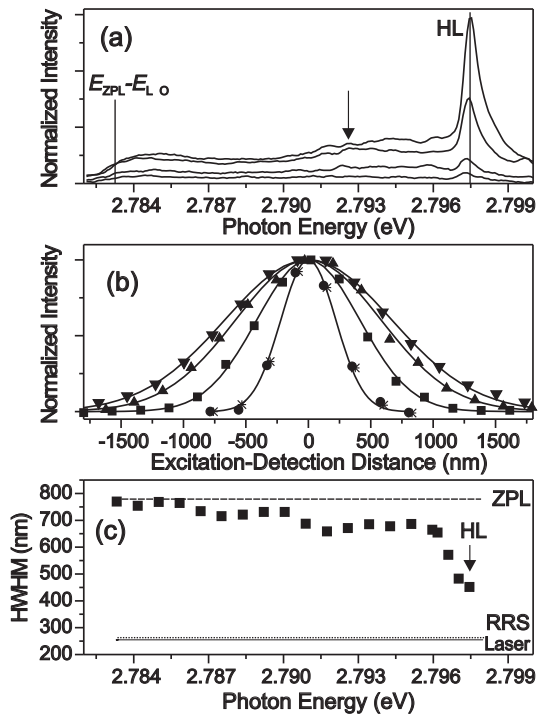}
 \caption{(a)~Spectra of PSB measured at different positions with respect to
 the excitation--laser spot. The distances from the excitation
 spot to the detection spot are (from top to bottom) 0, 460, 920 and
 1380~nm, respectively. The excitation laser energy is 2.8611~eV, with intensity of 1~$\mathrm{kW/cm^{2}}$.
 (b)~Spatial profiles of the excitation--laser spot (circles), RRS (stars), HL (squares)
 and ZPL (down--triangles). The spatial profile of one spectral
 component in the PSB, indicated as the arrow in (a), is also
 shown as the up--triangles. The solid lines represent the
 corresponding Gaussian fits to the data.
 (c)~HWHM of the spectral components in the PSB as function of the photon energy. The HWHMs are obtained
  by Gaussian fits of the spatial profiles of the selected components. The HWHMs of the ZPL (dashed), RRS (dots) and laser spot (solid) are
  also shown for comparison.
 }
\end{figure}
the excitation--detection distance, but is still visible at a
distance as large as 1380~nm. We also find a systematic change of
the PSB spectral shape, which reflects the relaxation of the hot
excitons during transport. These spectra allow us to obtain the
spatial profile of the HL peak, as shown by the squares in
Fig.~3(b). We also plot in the same figure the spatial profiles of
the laser spot (circles), the RRS peak (stars) and the ZPL
(down--triangles) measured in the same pinhole--scanning. The
profile of RRS peak is similar to that of the laser spot, while
the HL peak distribution is significantly wider. We note that the
difference in spatial extension can also be used as a method to
distinguish the HL and RRS processes. We also check the spatial
profile of a spectral component in the middle part of the PSB, as
indicated by the arrow in Fig.~3(a). The profile (up--triangles)
locates between the HL and ZPL profiles, showing the undergoing
relaxation. Such a feature is shown quantitatively in Fig.~3(c).
Decreasing the detection energy from HL peak to
$E_{\mathrm{ZPL}}-E_{\mathrm{LO}}$, i.e., scanning the arrow in
Fig.~3(a) from right to left within the PSB, we find an increase
of the HWHM (half--width at half--maxima, obtained by Gaussian
fits) approaching to that of ZPL. This confirms that the PSB
really monitors the exciton population as well as its spatial
extension.

As discussed before, the HL peak monitors the individual coherence
of the excitons. Thus the spatial profile of HL peak reflects the
coherent propagation of excitons in space. We note that HWHM of
the HL peak is related, but not equal, to the coherence length of
the excitons, since both the excitation spot and the detection
spot in our measurement are not given by delta functions in space.
Using Monte Carlo simulations for de--convolution, we deduce that
in our experimental setup the measured HWHM of the HL peak profile
is actually 1.35 times the coherence length.

Figure~4 shows the coherence lengths measured in this way for
several exciton kinetic energies by tuning the photon energy of
the excitation laser. Increasing the exciton kinetic energy we
find a slow drop of the coherence length. The decrease is less
than 20~\% as the kinetic energy is increased from 1.6 to
20.7~meV. We also show in the same figure (dots) the coherence
length calculated by only considering the acoustic phonon
scattering. For this estimation, the influence of static disorder
is not considered, thus the exciton is assumed to propagate
\begin{figure}
 \includegraphics[width=7cm]{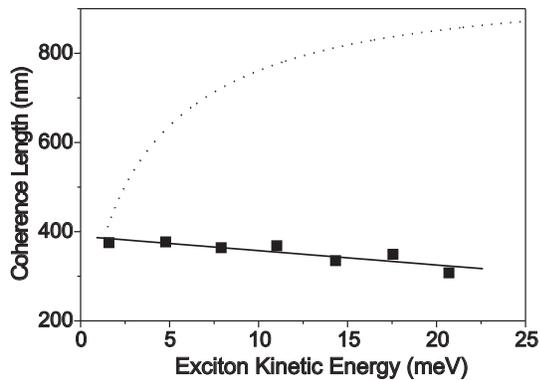}
 \caption{Exciton coherence lengths (squares) deduced from the spatial profiles of the HL peaks
 for several exciton kinetic energies. The solid line is a guide
 for the eyes. Also shown is the calculated coherence length (dots) limited by
 acoustic phonon scattering without considering the influence of
 disorder.
 }
\end{figure}
ballistically before the first acoustic--phonon scattering event.
The coherence length is thus calculated by simply multiplying the
group velocity obtained from a parabolic dispersion and the
acoustic--phonon scattering time\cite{b316552}. The discrepancy
observed in Fig.~4 suggests that the excitonic propagation is
limited by not only the acoustic--phonon scattering but also other
mechanisms like static disorder. The importance of the disorder on
excitonic properties of semiconductor quantum wells at low
temperature has been proven (see, e.g., Ref.~\cite{b4817149}).
Recently, theoretical investigation\cite{b608975} suggested a
strong influence of disorder on the spatiotemporal dynamics of
excitons in quantum wells under near--field pulsed excitation.
Strong disorder can eventually results in exciton localization.
Here, we are dealing with excitons with rather high kinetic energy
above the effective mobility edge. A detailed theoretical study on
the influence of static disorder on the coherence length of
excitons is beyond the scope of the present investigation.
However, Fig.~4 provides an important input for this kind of
studies. We also note that the coherence length determined from
the present investigation ($300\sim400$~nm) is about $25\sim30$
times the de Broglie wavelength of the exciton in ZnSe quantum
wells.

In summary, we show that the combination of a highly efficient,
sub--$\mu$m spatially resolved photoluminescence system with
phonon sideband spectroscopy enables one to investigate coherence
transport process in semiconductors. By monitoring the spatial
evolution of the exciton coherence, we directly determine the
exciton coherence length of $300\sim 400$~nm, which decreases
slowly when increasing the exciton kinetic energy.

We gratefully acknowledge the growth of excellent samples by the
group of M.~Heuken (RWTH Aachen) and useful discussion with
H.~Giessen (Universit\"at Bonn) and R.~von~Baltz (Universit\"at
Karlsruhe). This work was supported by the Deutsche
Forschungsgemeinschaft.

\end{document}